\documentclass[aip,pop,reprint,citeautoscript]{revtex4-2}

\usepackage{amsmath,amssymb}
\usepackage{ulem}

\usepackage{graphicx}
\usepackage{tabularx}
\usepackage{multirow}

\usepackage[per-mode=symbol]{siunitx}
\DeclareSIUnit\angstrom{\text{\AA}}
\DeclareSIUnit\au{\text{at.\,u.}}

\usepackage[colorlinks,urlcolor=blue,citecolor=blue]{hyperref}
\usepackage{color}
\usepackage[dvipsnames]{xcolor}

\begin{document}

\title{Real-time time-dependent density functional theory for high-energy density physics}

\author{Alina Kononov}
\email{akonono@sandia.gov}
\affiliation{Center for Computing Research, Sandia National Laboratories, Albuquerque NM 87123, USA \looseness=-1}

\author{Minh Nguyen}
\email{mnguye1@sandia.gov}
\affiliation{Center for Computing Research, Sandia National Laboratories, Albuquerque NM 87123, USA \looseness=-1}

\author{Andrew D. Baczewski}
\email{adbacze@sandia.gov}
\affiliation{Center for Computing Research, Sandia National Laboratories, Albuquerque NM 87123, USA \looseness=-1}

\begin{abstract}
Electronic response properties of high-energy density (HED) systems influence planetary structure, drive evolution of fusion targets, and underpin diagnostics in laboratory astrophysics.
Real-time time-dependent density functional theory (TDDFT) offers a versatile modeling framework capable of accurately predicting the dynamic response of HED materials --- including free-free, bound-free, and bound-bound contributions without requiring ad hoc state partitioning; capturing both collective and non-collective behavior; and applicable within the linear-response regime and beyond.
We review the theoretical formalism of real-time TDDFT as applied to HED systems, provide a practical tutorial for computing relevant response properties (dynamic structure factors, conductivity, and stopping power), and comment on avenues for further development of this powerful computational method in service of HED science.
\end{abstract}

\maketitle

\section{Introduction}
\label{sec:intro}

As a high-fidelity model of many-body electron dynamics, real-time time-dependent density functional theory (TDDFT) \cite{marques2012fundamentals,ullrich_time-dependent_2012,ullrich_time-dependent_2014} enables accurate predictions for a variety of electronic response properties crucial to high-energy density (HED) science. 
One such property is the dynamic structure factor (DSF) \cite{baczewski_x-ray_2016,white2025dynamical} governing x-ray Thomson scattering (XRTS), which is a leading experimental technique for inferring the temperature and density \cite{glenzer2007observations,kritcher2008ultrafast,fletcher2015ultrabright,descamps2020approach} of short-lived warm dense matter (WDM) samples generated at flagship HED science facilities \cite{nagler2015matter,zastrau2021high}.
Real-time TDDFT can also compute optical properties \cite{andrade2018negative,ramakrishna2023electrical,sharma2025group} like electrical conductivities entering hydrodynamic simulations of inertial confinement fusion \cite{haines2024charged}, pulsed-power experiments \cite{sinars2020review,stanek2024ethos}, and planetary dynamos \cite{stevenson2003planetary} or refractive indices influencing optical diagnostics in shock experiments \cite{kirsch2019refractive}.
The real-time formulation of TDDFT is perhaps most uniquely well-suited for precise calculations of electronic stopping power \cite{correa2018calculating,ding2018ab,magyar2016stopping,kononov2024reproducibility}, which underpins ion-driven fast-ignition \cite{roth2001fast} and alpha-particle self-heating \cite{zylstra_alpha-particle_2019} in fusion targets.

To compute these response properties, real-time TDDFT simulates the dynamics of quantum-mechanical electrons interacting with classical ions and classical external fields.
High-performance computing enables large calculations containing hundreds of ions and thousands of electrons, permitting realistic thermal disorder within the ionic subsystem. 
In principle, the treatment of electron-ion interactions can be arbitrarily precise, though high-Z systems become increasingly computationally intensive when strongly bound core electrons directly participate in the physical processes of interest.
Practical calculations also approximate electronic thermal statistics and quantum-mechanical effects within the mean-field treatment of electron-electron interactions.

With minimal approximations, TDDFT accurately captures a wide range of phenomenology relevant to WDM.
The extended systems possible with TDDFT support collective behavior paramount to free-electron response, while explicit ions also allow electron-ion scattering, bound-free processes, and bound-bound transitions.
All explicitly included electrons are modeled at the same level of theory, and all of these effects are generally simulated self-consistently.
As a non-perturbative approach, real-time TDDFT remains valid both within the linear-response regime and under strong perturbations.

The relative accuracy and favorable computational tradeoffs of TDDFT have made the method a gold standard for validating more efficient models of electronic response properties in the WDM regime \cite{grabowski2020review,stanek2024review}, especially since focused experiments \cite{zylstra:2015,malko2022proton,chen2021ultrafast,ofori2024dc} including model-free diagnostics \cite{dornheim2022accurate,gawne2025spectral} remain expensive, infrequent, and challenging.
For example, TDDFT calculations have guided the treatment of electron-ion collisions in dielectric models \cite{schorner2023x,hentschel2023improving,hentschel2025statistical} 
and justified the inclusion of a bound-bound term \cite{baczewski:2021} in the Chihara decomposition \cite{chihara2000interaction} for XRTS spectra.
Models that include more accurate treatments of electron-electron interactions and/or thermal effects \cite{brown2013path,dornheim2016ab,harsha2023thermofield,babbush2023quantum,rubin2024quantum} 
likewise offer opportunities to improve the approximations made within TDDFT.

Several implementations of real-time TDDFT are actively maintained and developed around the world, with some differences in available features and numerical algorithms.
For example, \textsc{SHRED} \cite{white2020fast,white2022mixed,sharma2023stochastic} includes stochastic, mixed stochastic-deterministic, and orbital-free variants, while \textsc{INQ} \cite{andrade2021inq} provides native GPU support and a flexible library paradigm.
A custom extension \cite{baczewski:2014,baczewski_x-ray_2016,magyar2016stopping} of \textsc{VASP} \cite{kresse:1996a,kresse:1996b,kresse:1999} inherits that code's particularly mature DFT implementation and well-validated PAW\cite{blochl_projector_1994} datasets, while \textsc{ELK} \cite{elk} and \textsc{EXCITING} \cite{gulans2014exciting,pela2021all} specialize in all-electron calculations.
A real-time TDDFT capability has also recently been implemented in Abinit \cite{gonze2002first,verstraete2025abinit}.
Other real-time TDDFT codes commonly used within the quantum chemistry and condensed matter physics communities include 
Octopus \cite{marques2003octopus,andrade2012time,tancogne2020octopus}, 
Qb@ll \cite{schleife2012plane,draeger2017massively}, 
NWChem \cite{valiev2010nwchem,apra2020nwchem}, 
and eQE \cite{krishtal2015subsystem}.

Here, we review general frameworks and practical considerations involved in TDDFT calculations for HED applications without dwelling upon the particularities of any individual code.
The remainder of this article is organized as follows.
Section \ref{sec:general} reviews the formalism of real-time TDDFT, including special considerations for HED conditions.
Then, we describe calculations within the linear-response regime to predict dynamic structure factors in Section \ref{sec:dsf} and optical properties including conductivity in Section \ref{sec:cond}.
Section \ref{sec:stopping} discusses calculations beyond linear response, focusing on electronic stopping power.
Finally, Section \ref{sec:conclusion} comments on the future outlook and remaining challenges for applying TDDFT to degenerate plasmas.
Except where otherwise indicated, we use atomic units where the electron mass, electron charge, and reduced Planck's constant are all unity ($m_e=e=\hbar=1$).

\section{General approach}
\label{sec:general}

Real-time TDDFT \cite{marques2012fundamentals,ullrich_time-dependent_2012,ullrich_time-dependent_2014} simulates electron dynamics in response to a time-dependent perturbation through the time-dependent Kohn-Sham equations:
\begin{equation}
    n(\mathbf{r},t) = \sum_j f_j(T)|\phi_j(\mathbf{r},t)|^2,
    \label{eq:tddensity}
\end{equation}
\begin{equation}
    i\frac{\partial}{\partial t}\phi_j(\mathbf{r},t) = H_\mathrm{KS}[n](\mathbf{r},t) \phi_j(\mathbf{r},t).
    \label{eq:tdks}
\end{equation}
The time-dependent electron density $n(\mathbf{r},t)$ is constructed from an auxiliary system of single-particle Kohn-Sham orbitals $\phi_j(\mathbf{r},t)$, with each single-particle probability density weighted by the corresponding Fermi occupation $f_j(T)$.
The Kohn-Sham orbitals evolve under a system of Schr\"odinger-like equations with an effective one-body Hamiltonian $H_\mathrm{KS}$ that is a functional of the electron density.
A plane-wave basis set is particularly amenable to representing Kohn-Sham orbitals in extended systems containing hundreds or thousands of electrons at elevated temperatures.

The Kohn-Sham Hamiltonian is given by
\begin{equation}
    H_\mathrm{KS}[n](\mathbf{r},t) = -\frac{\nabla^2}{2} + V_{\mathrm{KS}}[n](\mathbf{r},t),
    \label{eq:HKS}
\end{equation}
where the effective potential
\begin{equation}
    V_{\mathrm{KS}}[n](\mathbf{r},t) = V_{\mathrm{ext}}(\mathbf{r},t) + V_{\mathrm{H}}[n](\mathbf{r},t) + V_{\mathrm{XC}}[n](\mathbf{r},t)
    \label{eq:vks}
\end{equation}
includes contributions from electron-ion interactions and the external perturbation within $V_{\mathrm{ext}}$, the Hartree electron-electron interaction within $V_{\mathrm{H}}[n]$, and the exchange-correlation (XC) potential within $V_{\mathrm{XC}}[n]$.
$V_{\mathrm{H}}$ accounts for the classical Coulomb potential,
\begin{equation}
    V_{\mathrm{H}}[n](\mathbf{r},t) = \int \frac{n(\mathbf{r}',t)}{|\mathbf{r}-\mathbf{r}'|} d\mathbf{r}',
\end{equation}
while $V_{\mathrm{XC}}$ contains quantum-mechanical corrections to electron-electron interactions.

The formal foundation of TDDFT rests upon the Runge-Gross theorem \cite{runge_density-functional_1984}, which shows the existence of a one-to-one correspondence between $n(\mathbf{r},t)$ and $V_{\mathrm{KS}}[n](\mathbf{r},t)$ for a given set of initial Kohn-Sham orbitals $\phi_j(\mathbf{r},t=0)$.
Therefore, in principle TDDFT is capable of reproducing the exact dynamics of interacting electrons~\cite{van1999mapping}.
The exact effective potential can be computed from the exact time-dependent many-body density~\cite{maitra2016perspective}, but because this quantity is not generally known, $V_{\mathrm{XC}}$ must be approximated in practice.
Computational costs typically limit TDDFT calculations in the WDM regime to XC functionals that are local in time and local or semilocal in space, that is, the local density approximation (LDA) \cite{perdew_self-interaction_1981} or the generalized gradient approximation (GGA) \cite{perdew_generalized_1996}.
Recent advances include semilocal approximations with explicit temperature dependence \cite{karasiev_accurate_2014,groth_ab_initio_2017, karasiev_nonempirical_2018,kozlowski2023generalized} and efficient schemes for hybrid functionals within stochastic DFT formulations \cite{leveillee2025mixed}, but the influence of these improvements on TDDFT predictions for HED systems remains largely unexplored.

Another commonly approximated term is the external potential,
\begin{equation}
    V_{\mathrm{ext}}(\mathbf{r},t) = -\sum_I \frac{Z_I}{|\mathbf{r}-\mathbf{R}_I|} + V_{\mathrm{pert}}(\mathbf{r},t),
    \label{eq:vext}
\end{equation}
where $Z_I$ and $\mathbf{R}_I$ denote nuclear charges and positions and $V_{\mathrm{pert}}$ represents the perturbation.
High $Z_I$ present numerical challenges because tightly bound inner shell electrons require large basis sets and small time steps to resolve.
At moderate temperatures and densities, these core electrons remain fully occupied and nearly unaffected by nearby ions so that frozen atomic orbitals describe them effectively.
The pseudopotential approximation \cite{kleinman1982efficacious,vanderbilt_optimally_1985, hamann_optimized_2013} reduces the computational cost associated with inner shell electrons by incorporating them into $V_{\mathrm{ext}}$.
Alternatively, the projector augmented wave (PAW) method \cite{blochl_projector_1994} formally transforms the highly oscillatory all-electron wave functions into smoother pseudo-wave functions.

Pseudizing different sets of core electrons also allows isolating processes involving various core and valence electrons when feedback between their dynamics can be considered negligible.
For example, combining TDDFT results computed using multiple pseudization choices has enabled first-principles predictions of bound-free and free-free contributions to XRTS signals \cite{baczewski_x-ray_2016}.
Similarly, an efficient scheme for converging TDDFT stopping power calculations involves separately computing costly core contributions that are not sensitive to finite-size effects \cite{kononov_trajectory_2023}.

Crucially, TDDFT requires as input an appropriate ionic geometry determining $V_\mathrm{ext}$ and initial orbitals $\phi_j(\mathbf{r},t=0)$.
The ions may be arranged in a crystalline lattice representing solid material or thermalized to a desired temperature via \textit{ab intio} molecular dynamics.
To model HED materials, the initial orbitals are typically computed from Mermin DFT \cite{mermin1965thermal}, where Kohn-Sham occupations follow the Fermi distribution.
Notably, the ion and electron temperatures are free to differ, enabling modeling of e.g., isochorically heated materials.
While other initialization options aiming to capture realistic thermal fluctuations within the electronic subsystem have been explored \cite{modine2015representing}, unphysical electron thermalization dynamics \cite{kononov2022electron} pose a challenge for such approaches.

Given a Hamiltonian specification and a set of initial orbitals, a real-time TDDFT implementation evolves Eq.~\eqref{eq:tdks} using a discrete time step, which serves as an additional convergence parameter.
The numerical time-stepping algorithm influences the step size required to achieve a target precision, prompting concerted efforts to optimize the resulting cost-accuracy tradeoff \cite{castro2004propagators,rehn2019ode,kang2019pushing,kononov2022electron}.
While some early implementations favored explicit time-stepping algorithms like fourth-order Runge Kutta \cite{schleife2012plane}, implicit schemes like Crank-Nicolson \cite{crank1947practical} and enforced time-reversal symmetry \cite{castro2004propagators} are gaining popularity because of their superior charge conservation \cite{kononov2022electron,kang2019pushing,kononov2020pre} and particular suitability for the PAW method \cite{ojanpera_nonadiabatic_2012,baczewski:2014}.
An orthogonal version of PAW also allows for its efficient use in explicit time-propagation methods \cite{li2020real,nguyen2024time}.

Static DFT calculations typically take advantage of reciprocal-space symmetries in order to reduce computational cost.
However, such reductions must be carefully re-examined in the time-dependent case: the perturbing potential can break symmetries, and artificially enforcing them can lead to incorrect results.
TDDFT calculations often avoid this issue when thermal disorder and/or physical processes of interest require large supercells, leading to a very small Brillouin zone adequately sampled by the $\Gamma$ point.
However, the crystal structure of ambient or isochorically heated material could be modeled with a primitive unit cell, in which case issues of reciprocal-space sampling are more important.

\section{Dynamic structure factor}
\label{sec:dsf}

The dynamic structure factor (DSF) can be extracted from real-time TDDFT calculations by analyzing the density response to a weak sinusoidal potential as first described in Ref.~\onlinecite{sakko2010time} and adapted to HED conditions in Ref.~\onlinecite{baczewski_x-ray_2016}.
Within the linear regime, 
the density response $\delta n(\mathbf{r},t) = n(\mathbf{r},t) - n(\mathbf{r},0)$ is related to the perturbing potential $V_{\mathrm{pert}}(\mathbf{r},t) = V_{\mathrm{ext}}(\mathbf{r},t) - V_{\mathrm{ext}}(\mathbf{r},0)$ via the linear density-density response function $\chi_{nn}(\mathbf{r},\mathbf{r}',t)$:
\begin{equation}
    \delta n(\mathbf{r},t) = \int_0^\infty d\tau \int_\Omega d\mathbf{r}'\, \chi_{nn}(\mathbf{r},\mathbf{r}',\tau) V_{\mathrm{pert}}(\mathbf{r}',t-\tau),
    \label{eq:density_response}
\end{equation}
where $\Omega$ denotes the volume of the supercell and $\tau$ is a delay between the perturbation and response.
The Fourier transform of this relationship takes the form
\begin{equation}
    \delta \tilde{n}(\mathbf{q},\omega) = \sum_{\mathbf{q}'}\, \tilde{\chi}_{nn}(\mathbf{q},-\mathbf{q}',\omega) \tilde{V}_{\mathrm{pert}}(\mathbf{q}',\omega),
    \label{eq:tilde_n_sum}
\end{equation}
where $\mathbf{q}$, $\mathbf{q}'$ are reciprocal space vectors and the Fourier transforms are defined such that
\begin{align}
    \tilde{\chi}_{nn}&(\mathbf{q},-\mathbf{q}',\omega)=\\
    & \int d\mathbf{r} \int d\mathbf{r}' \int d\tau \, \chi_{nn}(\mathbf{r},\mathbf{r}',\tau) e^{i\omega \tau} e^{i\mathbf{q}\cdot\mathbf{r}} e^{-i\mathbf{q}'\cdot\mathbf{r}'}.
\end{align}

Macroscopic response properties can be derived from this microscopic description by considering the potential induced by the density response,
\begin{equation}
    \tilde{V}_\mathrm{ind}(\mathbf{q},\omega) = \frac{4\pi}{|\mathbf{q}|^2} \delta \tilde{n}(\mathbf{q},\omega),
\end{equation}
so that the sum of the perturbing and induced potentials is
\begin{align}
    \tilde{V}_\mathrm{tot}&(\mathbf{q},\omega) 
    = \tilde{V}_\mathrm{pert}(\mathbf{q},\omega) + \tilde{V}_\mathrm{ind}(\mathbf{q},\omega)\\
    &= \sum_{\mathbf{q}'}\, \left(\delta_{\mathbf{q},\mathbf{q}'} + \frac{4\pi}{|\mathbf{q}|^2}\tilde{\chi}_{nn}(\mathbf{q},-\mathbf{q}',\omega)\right) \tilde{V}_{\mathrm{pert}}(\mathbf{q}',\omega).
    \label{eq:lrVtot}
\end{align}
Therefore, the macroscopic dielectric function can be computed from
\begin{align}
    \frac{1}{\varepsilon(\mathbf{q},\omega)} 
    =& 1 + \frac{4\pi}{|\mathbf{q}|^2} \chi_{nn}(\mathbf{q}, -\mathbf{q}, \omega),
    \label{eq:dielectric}
\end{align}
where only the $\mathbf{q}=\mathbf{q}'$ term of Eq.~\eqref{eq:lrVtot} remains after relating the macroscopic and microscopic response functions \cite{onida2002electronic}.

All other linear-response functions fundamentally relate to $\varepsilon(\mathbf{q},\omega)$.
In particular, the DSF is given by
\begin{equation}
    S(\mathbf{q}, \omega) = -\frac{1}{\pi n_i}\frac{\mathrm{Im}\left[ \tilde{\chi}_{nn}(\mathbf{q},-\mathbf{q}, \omega) \right]}{1-e^{-\omega/(k_B T)}},
    \label{eq:dsf}
\end{equation}
where $n_i$ is the ion density and $k_B$ is Boltzmann's constant.

In principle, any perturbing potential that is compatible with the periodicity of the system and has support over the desired energy and momentum transfers (i.e., $\tilde{V}_\mathrm{pert}(\mathbf{q},\omega)\neq 0$ for the $\mathbf{q},\omega$ of interest) could be applied.
Naively, it would seem particularly convenient to include only a single wavevector $\mathbf{q}$ corresponding to the scattering angle of interest, i.e., apply
\begin{equation}    
    V_\mathrm{pert}(\mathbf{r},t) = I_0 e^{i \mathbf{q}\cdot\mathbf{r}} f(t)
    \label{eq:complexVpert}
\end{equation}
where $I_0$ is the perturbation intensity and $f(t)$ is a temporal envelope function satisfying $\int_0^\infty f(t) = 1$.
Then, Eq.~\eqref{eq:tilde_n_sum} would appear to simplify so that the needed component of the density-density response function is
\begin{equation}
     \tilde{\chi}_{nn}(\mathbf{q},-\mathbf{q},\omega) = \frac{\delta \tilde{n}(\mathbf{q},\omega)}{\tilde{V}_{\mathrm{pert}}(\mathbf{q},\omega)}.
\end{equation}

However, since a physical potential must be Hermitian, in practice, two calculations are needed with sinusoidal perturbing potentials
\begin{align}
\label{eq:cosVpert}
V_\mathrm{pert}^{\mathrm{C}}(\mathbf{r},t) =& I_0 f(t)\cos(\mathbf{q}\cdot\mathbf{r}), \\ 
\label{eq:sinVpert}
V_\mathrm{pert}^{\mathrm{S}}(\mathbf{r},t) =& I_0 f(t)\sin(\mathbf{q}\cdot\mathbf{r}). 
\end{align}
Under these perturbations, Eq.~\eqref{eq:tilde_n_sum} reduces to
\begin{align}
        \delta \tilde{n}^{\mathrm{C}}(\mathbf{q},\omega) =& \frac{V_0 \tilde{f}(\omega)}{2} \left( \tilde{\chi}_{nn}(\mathbf{q},-\mathbf{q},\omega) + \tilde{\chi}_{nn}(\mathbf{q},\mathbf{q},\omega) \right),\\
        \delta \tilde{n}^{\mathrm{S}}(\mathbf{q},\omega) =& \frac{V_0 \tilde{f}(\omega)}{2i} \left( \tilde{\chi}_{nn}(\mathbf{q},-\mathbf{q},\omega) - \tilde{\chi}_{nn}(\mathbf{q},\mathbf{q},\omega) \right),
\end{align}
respectively.
Combining the two responses by computing
\begin{align}
    \delta \tilde{n}^\mathrm{T}(\mathbf{q},\omega) &= 
    \delta \tilde{n}^{\mathrm{C}}(\mathbf{q},\omega) + i \delta \tilde{n}^{\mathrm{S}}(\mathbf{q},\omega) 
    \label{eq:sincos_response1} \\
    &= I_0 \tilde{f}(\omega) \tilde{\chi}_{nn}(\mathbf{q},-\mathbf{q},\omega)
    \label{eq:sincos_response2}
\end{align}
then enables isolating $\tilde{\chi}_{nn}(\mathbf{q},-\mathbf{q},\omega)$ to evaluate Eq.~\eqref{eq:dsf}.

Notably, directly applying the complex potential of Eq.~\eqref{eq:complexVpert} within a single TDDFT simulation is not equivalent to the split potential framework of Eqs.~\eqref{eq:cosVpert}\,--\,\eqref{eq:sincos_response2}.
A generic non-Hermitian Hamiltonian could generate dissipative dynamics violating charge conservation.
Moreover, the density response $\delta n(\mathbf{r},t)$ is always real-valued regardless of the complex phase of $V_\mathrm{pert}(\mathbf{r},t)$.
We emphasize that the linear-response formalism starting from Eq.~\eqref{eq:density_response} only holds for Hermitian potentials.

\begin{figure}
    \centering
    
    \includegraphics{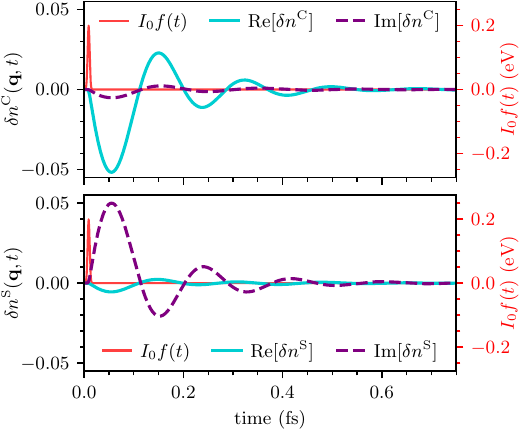}
    \caption{Exemplary pair of TDDFT density response calculations for a particular choice of $\mathbf{q}$.
    In the top and bottom panels, the perturbing potential takes the form of Eq.~\eqref{eq:cosVpert} and \eqref{eq:sinVpert}, respectively, where the temporal envelope $I_0 f(t)$ is shown in red.
    The corresponding real and imaginary parts of the density response computed through real-time evolution of Eq.~\eqref{eq:tdks} are shown in teal and purple, respectively.
    }
    \label{fig:dsf_example_t}
\end{figure}

Fig.~\ref{fig:dsf_example_t} illustrates an example of two TDDFT calculations used to predict the DSF of solid-density aluminum at a temperature of \SI{1}{\electronvolt} and momentum transfer of $q=\SI{1.55}{\angstrom}^{-1}$ in Refs.~\onlinecite{hentschel2023improving,hentschel2025statistical}.
The perturbing potential is a short Gaussian pulse of width $t_w=\SI{2}{\atto\second}$ centered at $t_d=\SI{10}{\atto\second}$,
\begin{equation}
    f_{\mathrm{G}}(t) = \frac{1}{\sqrt{2\pi}t_w} e^{-(t-t_d)^2/(2t_w^2)},
    \label{eq:gauss_pulse}
\end{equation}
modulated by a sinusoidal spatial dependence (see Eqs.~\eqref{eq:cosVpert} and \eqref{eq:sinVpert}).
As the orbitals evolve in response, the electron density shows clear plasmonic oscillations.

\begin{figure}
    \centering
    
    \includegraphics{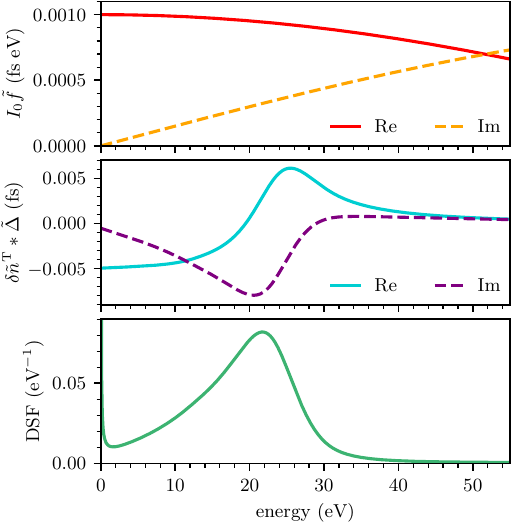}
    \caption{Exemplary post-processing steps to compute the DSF from the time-dependent density-response data shown in Fig.~\ref{fig:dsf_example_t}.
    First, the perturbing potential and total density response $\delta n^\mathrm{T}(\mathbf{q},t) = \delta n^\mathrm{C}(\mathbf{q},t) + i\delta n^\mathrm{S}(\mathbf{q},t)$ are transformed into the frequency domain (top two panels) after scaling by a window function $\Delta(t)$ in the latter case.
    Then, the density-density response function and corresponding DSF are evaluated according to Eqs.~\eqref{eq:sincos_response2} and \eqref{eq:dsf}.
    }
    \label{fig:dsf_example_w}
\end{figure}

Fig.~\ref{fig:dsf_example_w} illustrates the subsequent analysis involved in extracting the DSF.
For a sufficiently short perturbing pulse, the Fourier transform
\begin{equation}
    \tilde{f}_{\mathrm{G}}(\omega) = e^{i\omega t_d}\, e^{-(\omega t_w)^2/2}
    \label{eq:gauss_FT}
\end{equation}
maintains strong support over a wide range of frequencies.
To dampen numerical artifacts arising from the finite-duration of the real-time signal,
we scale the real-time density response by a Gaussian window function
\begin{equation}
    \Delta(t) = e^{-(\gamma t)^2}
    \label{eq:gaussian_broadening}
\end{equation}
of width given by $\hbar\gamma = \SI{1}{\electronvolt}$.
As a consequence of the convolution theorem, this step is similar to applying a Gaussian broadening to the resulting spectrum.
Finally, dividing the density response by the perturbation in the frequency domain gives the density-density response function and DSF (see Eqs.~\eqref{eq:sincos_response2} and \eqref{eq:dsf}).

In order to obey periodic boundary conditions, $\mathbf{q}$ must be an integer linear combination of the reciprocal lattice vectors associated with the simulation cell.
For example, a cubic cell of length $L$ can support $\mathbf{q}$ vectors of the form $\frac{2\pi}{L}(n_x \hat{x} + n_y \hat{y} + n_z \hat{z})$, where $n_x,n_y,n_z \in \mathbb{Z}$.
Despite this discrete constraint, arbitrary $\mathbf{q}$ magnitudes can be accessed for melted systems simply by adjusting the dimensions of an orthorhombic simulation cell.
However, accessing very small momentum transfers can require very large simulation cells, and we return to this topic in Section \ref{sec:cond}.

\begin{figure}
    \centering
    
    \includegraphics{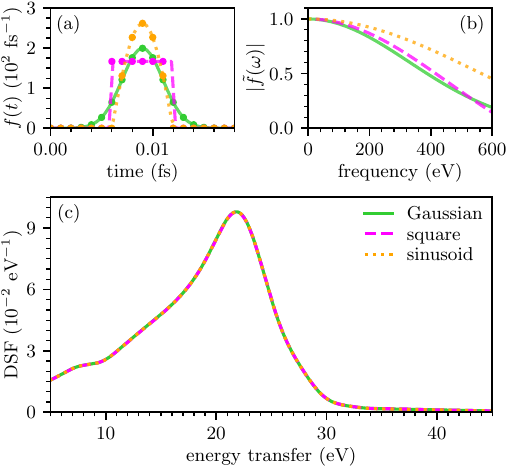}
    \caption{Robustness of DSF predictions across three different types of perturbation envelope $f(t)$.
    Panels (a) and (b) show the Gaussian (solid green), square (dashed pink), and sinusoidal (dotted orange) pulses in the time and frequency domain, respectively, where points indicate the discrete time steps.
    Panel (c) shows excellent agreement among the resulting DSFs.
    }
    \label{fig:dsf_probes}
\end{figure}

Beyond maintaining support over the frequency range of interest, the specific shape of the perturbation's temporal envelope $f(t)$ is not important.
Besides the smooth Gaussian form of Eq.~\eqref{eq:gauss_pulse}, other options like a square or sinusoidal pulse are also possible.
Within the linear regime, the density response $\delta\tilde{n}(\mathbf{q},\omega)$ is proportional to $\delta\tilde{f}(\omega)$ such that the dependence on $f(t)$ cancels out upon computing $\tilde{\chi}_{nn}(\mathbf{q},\mathbf{-q},\omega)$.
Fig.~\ref{fig:dsf_probes} demonstrates that the predicted DSF is not sensitive to details of the perturbation.
In particular, no numerical artifacts arise from discontinuities in $f(t)$ or its derivative.

Importantly, the perturbation intensity $I_0$ must be small enough to stay in the linear regime.
For a large $I_0$, the density response in Eq.~\eqref{eq:density_response} would also contain non-negligible contributions from higher-order response functions.
In practice, multiple tests with different values of $I_0$ are required to verify that the density response is indeed proportional to the magnitude of the perturbation, i.e., that the perturbation is weak enough that linear response dominates the electron dynamics.

However, under weaker perturbations, numerical errors may become more noticeable relative to the physical response.
In particular, the linear-response framework assumes the initial states remain stationary, i.e., $n(\mathbf{r},t)=n(\mathbf{r},0)$ under no perturbation.
Exact unperturbed evolution of exact Mermin-Kohn-Sham states would indeed produce a constant density, with each state simply rotating by a complex phase given by the corresponding eigenvalue.
However, numerical errors in the initial condition or subsequent time propagation may disturb stationarity and pollute the apparent density response $\delta n(\mathbf{r},t)$.
Thus, initial states for real-time TDDFT may require a stricter convergence criterion than a Mermin-DFT calculation of static electronic properties would typically demand.
In cases where the errors remain within the linear-response regime but further improving numerical precision proves impractical, correcting the computed response by subtracting the fictitious dynamics of the unperturbed system can 
restore accuracy.

\begin{figure}
    \centering
    \includegraphics{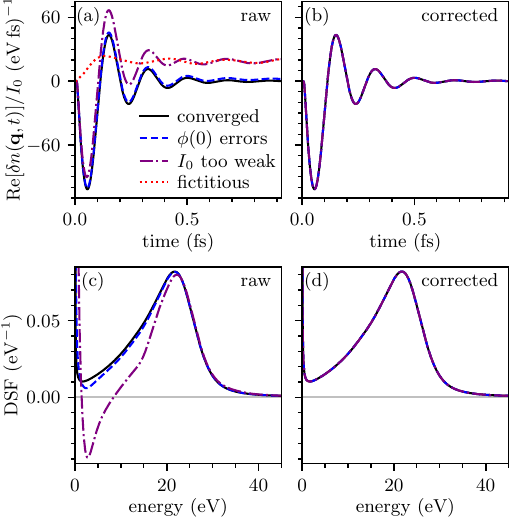}
    \caption{
    Propagation of initial-state errors in real-time TDDFT calculations of linear-response functions.
    Panels (a) and (b) show the real part of the density response $\delta n(\mathbf{q},t)$ normalized by the perturbation intensity $I_0$, 
    while panels (c) and (d) show the resulting dynamic structure factor.
    Fully converged results (solid black) are compared to cases where the initial states $\phi_j(\mathbf{r},t=0)$ were less converged (dashed blue and dashed-dotted purple).
    When $I_0$ is too weak compared to the initial-state errors (dashed-dotted purple), fictitious dynamics (dotted red) pollute the density response in (a) and distort the resulting DSF in (c).
    For sufficiently small initial-state errors, subtracting the fictitious dynamics from the computed response can recover accurate results as shown in (b) and (d).
}
    
    \label{fig:lr_intrinsic}
\end{figure}

Fig.~\ref{fig:lr_intrinsic} shows an example where fictitious dynamics arising from small errors in the initial states distort DSF predictions for solid-density aluminum at a temperature of \SI{1}{\electronvolt} and wavevector $|\mathbf{q}|=1.55$\SI{}{\angstrom}$^{-1}$.
Within the linear regime, the density response should scale with the perturbation magnitude, and the response functions should remain independent of perturbation magnitude.
However, in the presence of initial-state errors, the fictitious dynamics increasingly dominates the density response under weak perturbations, spoiling the expected linear-response behavior.
In this case, the initial-state errors manifest as a negative bias in the low-frequency DSF, which can lead to unphysical results violating causality.
After subtracting the fictitious dynamics of the unperturbed system from the computed density response, the predicted response function becomes independent of perturbation strength, as expected, and recovers the fully converged result.

Similarly, careful testing is required to ensure satisfactory convergence of the spectral features of interest with respect to the numerical time step, electronic basis size, and post-processing choices.
While the plasmon peak tends to be fairly robust, we find that the $\omega\rightarrow 0$ limit and high-energy tail ($\omega \gtrsim \SI{200}{\electronvolt}$) can be very sensitive to computational parameters and post-processing choices.
Thus, the sum rule
\begin{equation}
    \int_{-\infty}^{\infty} d\omega\, \omega S(\mathbf{q},\omega) = \frac{1}{2}N |\mathbf{q}|^2,
    \label{eq:dsffsum}
\end{equation}
where $N$ is the total number of explicitly simulated electrons per ion,
is generally difficult to satisfy precisely \cite{baczewski_x-ray_2016}.

Evaluating the DSF according to the form in Eq.~\eqref{eq:dsf} requires regularization near $\omega=0$ because of the delta-function-like behavior of the elastic peak.
When the physics of interest is entirely in the inelastic response (e.g., the plasmon dispersion), this complication presents no problems because the relevant parts of the DSF are away from $\omega=0$.
But when comparing the DSF from TDDFT to experimental XRTS measurements (e.g., convolving it with an instrument function, as in Fig. 13 in Ref.~\onlinecite{dornheim2023electronic}), one can use a regularized DSF of the form
\begin{equation}
    S(\mathbf{q},\omega\approx 0) = S_\mathrm{ee}(\mathbf{q}) \delta(\omega)+S_\mathrm{reg}(\mathbf{q},\omega),\label{eq:dsf_reg}
\end{equation}
where $S_\mathrm{ee}(\mathbf{q})$ is the static structure factor and $S_\mathrm{reg}(\mathbf{q},\omega)$ is a version of the DSF in which pathological behavior near $\omega\rightarrow0$ is softened.

The nature of these $\omega\rightarrow0$ pathologies becomes evident when Taylor expanding the numerator and denominator in Eq.~\eqref{eq:dsf} about $\omega=0$,
\begin{equation}
    S(\mathbf{q}, \omega) \approx  -\frac{k_B T}{\pi n_i}\frac{\mathrm{Im}\left[\tilde{\chi}_{nn}(0)+\left.\omega\frac{\partial\tilde{\chi}_{nn}}{\partial \omega}\right|_{\omega=0}\right]+ \mathcal{O}(\omega^2)}{\omega + \mathcal{O}(\omega^2)},
\end{equation}
where we have dropped the $\mathbf{q}$-dependence of $\tilde{\chi}_{nn}$ for brevity.
In particular, any non-zero imaginary contribution to $\tilde{\chi}_{nn}(\mathbf{q},-\mathbf{q}, 0)$ will lead to a $1/\omega$ divergence in the computed DSF.
While these contributions near $\omega=0$ could be physical (albeit subject to both numerical and systematic errors), they are slow to average over snapshots because of their amplification by this $1/\omega$ factor.
So, in postprocessing the density-density response function from a single snapshot near $\omega=0$, one can simply remove the DC component, leaving
\begin{equation}
    S_\mathrm{reg}(\mathbf{q},\omega \approx 0)=-\frac{k_BT}{\pi n_i}\mathrm{Im}\left[\frac{\partial\tilde{\chi}_{nn}}{\partial \omega}(\mathbf{q},-\mathbf{q}, 0)\right].
\end{equation}
It is then straightforward to evaluate the regularized form of the DSF in Eq.~\eqref{eq:dsf_reg} for these very small values of $\omega$.
 
\section{Conductivity and optical properties}
\label{sec:cond}

Predicting optical properties using the density-density response formalism described in Eqs.~\eqref{eq:density_response}\,--\,\eqref{eq:dielectric} would be challenging.
Since unitary time evolution conserves charge, $\delta\tilde{n}(\mathbf{q},\omega)$ and $\tilde{\chi}_{nn}(\mathbf{q},-\mathbf{q},\omega)$ vanish at $\mathbf{q}=0$.
In principle, Eq.~\eqref{eq:dielectric} could still be evaluated in the small $\mathbf{q}$ limit, but such an approach may require impractically large simulation cells.

\begin{figure}
    \centering
    
    \includegraphics{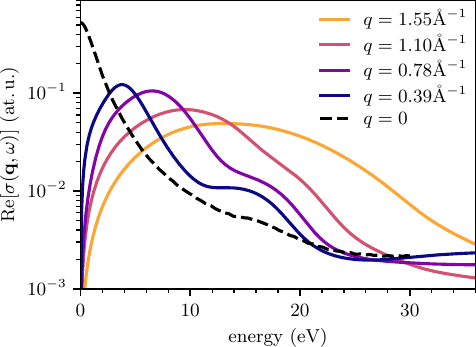}
    \caption{Low-$q$ response functions predicted using real-time TDDFT for solid-density aluminum at a temperature of \SI{1}{\electronvolt}.
    The solid curves were computed through the density-response formalism of Section~\ref{sec:dsf}, while the dashed curve at \mbox{$\mathbf{q}=0$} was computed through the current-response formalism of Section~\ref{sec:cond}.
    }
    \label{fig:lowq}
\end{figure}

For example, Fig.~\ref{fig:lowq} depicts the real part of the complex conductivity,
\begin{align}
    \sigma(\mathbf{q},\omega) =& \frac{i\omega}{4\pi} (1-\varepsilon(\mathbf{q},\omega))
    \label{eq:cond_dielectric} \\
    =& \frac{i\omega}{4\pi} \left(1- \frac{1}{1 + \frac{4\pi}{|\mathbf{q}|^2} \tilde{\chi}_{nn}(\mathbf{q}, -\mathbf{q}, \omega)} \right),
    \label{eq:cond_chinn}
\end{align}
as computed from the density-density response function at small $\mathbf{q}$ according to the approach of Section \ref{sec:dsf}.
While relatively small 32-atom simulation cells suffice for the $|\mathbf{q}|\geq \SI{0.78}{\angstrom^{-1}}$ cases, the smallest non-zero value of $|\mathbf{q}|$ was obtained with an elongated 64-atom cell where $\mathbf{q}$ lies parallel to the long axis of length \SI{16.2}{\angstrom}.
Slow convergence to the $\mathbf{q}\rightarrow 0$ limit, along with the inverse proportionality between minimum accessible $|\mathbf{q}|$ and simulation cell dimensions, would require even larger cells to approach optical properties.
Similar slow convergence behavior may also challenge experimental proposals to infer optical properties from low-$\mathbf{q}$ XRTS spectra \cite{gawne2024ultrahigh}.

Simultaneously accessing the $\mathbf{q}\rightarrow 0$ and $\omega\rightarrow 0$ limits to predict DC conductivity from the density-response framework presents further difficulties.
Since the computed $\tilde{\chi}_{nn}$ does not correctly approach the real axis for small $\omega$ (see discussion of the $\omega\rightarrow 0$ DSF divergence in Section \ref{sec:dsf}), $\mathrm{Im}[\varepsilon(\mathbf{q},\omega)]$ does not diverge appropriately as $\omega\rightarrow 0$ and therefore the resulting $\mathrm{Re}[\sigma(\mathbf{q},0)]$ always vanishes (see Fig.~\ref{fig:lowq}).
These limitations of the density-response framework necessitate an alternative approach for predicting optical properties like DC conductivity from real-time TDDFT.

As a formally equivalent but more computationally convenient approach \cite{sangalli2017optical},
optical properties can be computed from time-dependent \textit{current} density functional theory (TD-CDFT), which introduces an external vector potential $\mathbf{A}_\mathrm{ext}(\mathbf{r},t)$ within the kinetic energy term of the Hamiltonian:
\begin{equation}
    H_\mathrm{KS}[n](\mathbf{r},t) = \frac{1}{2}\left(i\nabla + \mathbf{A}_\mathrm{ext}(\mathbf{r},t) \right)^2 + V_{\mathrm{KS}}[n](\mathbf{r},t).
    \label{eq:HKSA}
\end{equation}
Rather than the density response, the observable of interest becomes the current density
\begin{align}
    \mathbf{j}(\mathbf{r},t) = -&\sum_j f_j(T) \mathrm{Im}\left[ \phi_j^*(\mathbf{r},t) \nabla \phi_j(\mathbf{r},t)  \right] \nonumber\\
    +& \mathbf{A}_\mathrm{ext}(\mathbf{r},t) n(\mathbf{r},t) .
    \label{eq:current}
\end{align}
Similar to the Runge-Gross theorem \cite{runge_density-functional_1984} underlying TD-DFT, for a fixed initial state, the mapping between the external potentials $V_\mathrm{ext}(\mathbf{r},t)$, $\mathbf{A}_\mathrm{ext}(\mathbf{r},t)$ and the densities $n(\mathbf{r},t)$, $j(\mathbf{r},t)$ is unique up to gauge transformations \cite{vignale2004mapping}.

Optical properties may be obtained by imposing a perturbation corresponding to a uniform electric field
\begin{equation}
    \mathbf{E} = -\nabla V_\mathrm{pert} - \frac{\partial \mathbf{A}_\mathrm{ext}}{\partial t}.
    \label{eq:Efield}
\end{equation}
In periodic systems, the velocity gauge where $V_\mathrm{pert}=0$ and $\mathbf{E} = - \partial \mathbf{A}_\mathrm{ext}/\partial t$ is particularly convenient.
Through Ohm's law, the complex conductivity tensor $\overleftrightarrow{\sigma}$ relates the current density and the perturbing electric field in the frequency domain:
\begin{equation}
    \tilde{j}_\alpha(\omega) = \sum_\beta \sigma_{\alpha\beta}(\omega) \tilde{E}_\beta(\omega),
    \label{eq:ohm}
\end{equation}
where $\alpha$, $\beta$ denote Cartesian indices.

This TD-CDFT approach overcomes the $q\rightarrow 0$ limitations of Eq.~\eqref{eq:cond_chinn} through its underlying reliance on the current-current response tensor $\overleftrightarrow{\chi}_{jj}$ rather than the density-density response function.
Analogously to Eq.~\eqref{eq:density_response}, $\overleftrightarrow{\chi}_{jj}$ relates the current response $\delta j(\mathbf{r},t) = j(\mathbf{r},t) - j(\mathbf{r},0)$ to the external vector potential:
\begin{equation}
    \delta \mathbf{j}(\mathbf{r},t) = \int_0^\infty d\tau \int_\Omega d\mathbf{r}'\, \overleftrightarrow{\chi}_{jj}(\mathbf{r},\mathbf{r}',\tau) \mathbf{A}_\mathrm{ext}(\mathbf{r}',t-\tau),
    \label{eq:current_response}
\end{equation}
\begin{equation}
    \delta \tilde{\mathbf{j}}(\mathbf{q},\omega) = \sum_{\mathbf{q}'}\, \tilde{\overleftrightarrow{\chi}}_{jj}(\mathbf{q},-\mathbf{q}',\omega) \tilde{\mathbf{A}}_{\mathrm{ext}}(\mathbf{q}',\omega).
    \label{eq:tilde_j_sum}
\end{equation}
Using $\mathbf{E}(\mathbf{q},\omega)=-i\omega\mathbf{A}_\mathrm{ext}(\mathbf{q},\omega)$ and again discarding $\mathbf{q}\neq\mathbf{q}'$ terms to access macroscopic response, we thus have
\begin{equation}
    \overleftrightarrow{\sigma}(\mathbf{q},\omega) = \frac{i}{\omega} \tilde{\overleftrightarrow{\chi}}_{jj}(\mathbf{q},-\mathbf{q},\omega),
\end{equation}
which poses no immediate difficulties at $\mathbf{q}=0$.
The density-density and current-current response functions are related through the continuity equation $\mathbf{q}\cdot\delta\tilde{\mathbf{j}} = \omega \delta\tilde{n}$, which requires \cite{sangalli2017optical}
\begin{equation}
    \omega^2 \tilde{\chi}_{nn} = \mathbf{q}\cdot(\overleftrightarrow{\chi}_{jj} \mathbf{q}).
\end{equation}
Notably, the density-density framework only provides information about the longitudinal response where the perturbing fields are parallel to $\mathbf{q}$, whereas the current-current framework can also access transverse components of the response.

In calculating conductivity in the linear-response regime using real-time TD-CDFT, it can suffice to weakly perturb the system in one direction via a delta kick, e.g., $\mathbf{E}(t)=E_0\,\delta(t)\,\hat{\mathbf{x}}$, and only consider the corresponding diagonal element of the conductivity tensor $\sigma_{xx}$ \cite{ramakrishna2023electrical}.
Under this type of perturbation, Eq.~\eqref{eq:ohm} reduces to
\begin{equation}
    \sigma_{xx}(\omega) = \frac{\tilde{j}_x(\omega)}{E_0}.
    \label{eq:conductivity_lr}
\end{equation}
This kick corresponds to a vector potential of the form
\begin{equation}
    \label{eq:kick vector potential}
    \mathbf{A}_\mathrm{ext}(t) = -E_0\Theta(t) \hat{\mathbf{x}},
\end{equation}
where $\Theta(t)$ is the Heaviside function centered at $t=0$.
The advantage of this type of perturbation is that the external field contains all frequencies with the same magnitude and naturally separates the external field at $t=0$ and induced fields at $t>0$.
Similar to the discussion of Fig.~\ref{fig:dsf_probes}, conductivity results should not depend on the specific shape of the temporal pulse.
In anisotropic solids or melted cases with significant sensitivities to atomic configuration, explicitly computing each diagonal element of the conductivity tensor and taking their average may be preferable.

After calculating the time-dependent current density via Eq.~\eqref{eq:current} for a finite amount of time, a Fourier transform gives $\tilde{\mathbf{j}}(\omega)$.
Similar to the density response post-processing steps discussed in Section \ref{sec:dsf} (see Eq.~\eqref{eq:gaussian_broadening} and accompanying text),
the time-dependent current density must be scaled by a window function $\Delta(t)$ such as a Gaussian or decaying exponential to mitigate spectral leakage arising from the finite duration of the signal. 
Here, we consider the Gaussian window function of Eq.~\eqref{eq:gaussian_broadening}, which also effectively applies a Gaussian broadening of width $\gamma$ to the dynamic conductivity.

For conducting systems without sharp spectral features, we find that the dynamic conductivity at frequencies away from the DC limit ($\omega\gtrsim\SI{3}{\electronvolt}$) is generally insensitive to details of the window function. 
However, $\gamma$ has a greater influence near the DC limit due to the presence of a non-trivial, unphysical long-time average around which the current density fluctuates. 
For small $\gamma$, the unreliable long-time behavior of the current density increasingly biases the low-frequency limit of the conductivity, given by
\begin{equation}
    \sigma(0) = \frac{1}{E_0} \int \Delta(t) j(t) dt .
\end{equation}
The DC conductivity for melted systems represented by large supercells shows less sensitivity to windowing than for solids with crystalline order, as the former exhibit smaller magnitude long-time average currents than the latter. 

\begin{figure}[h]
    \centering
    \includegraphics{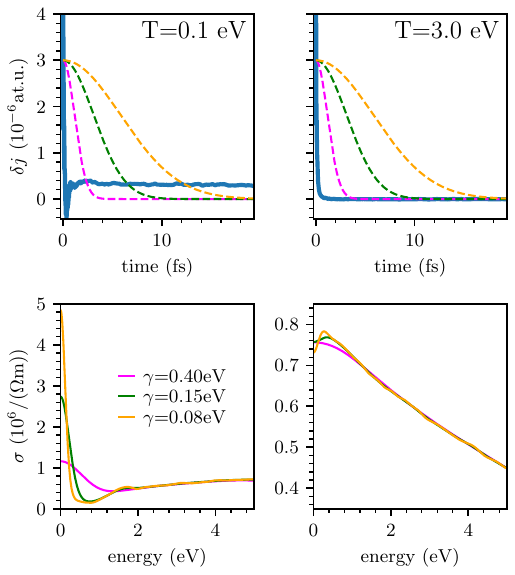}
    \caption{Exemplary real-time TDDFT conductivity calculations using Eq.~\eqref{eq:conductivity_lr} for beryllium at a density of \SI{1.97}{\gram\per\centi\meter\cubed}. 
    The left and right columns correspond to temperatures of \SI{0.1}{\electronvolt} and \SI{3.0}{\electronvolt}, respectively.
    The top row shows time-dependent macroscopic current density response overlaid with scaled Gaussian window functions of various widths as defined in Eq.~\eqref{eq:gaussian_broadening}. 
    The bottom row shows dynamic conductivities calculated using those window functions.
    The simulations used supercells containing 128 beryllium atoms.    
    }
    \label{fig:conductivity_calc}
\end{figure}

Fig.~\ref{fig:conductivity_calc} highlights this behavior with TDDFT conductivity calculations for beryllium at a density of \SI{1.97}{\gram\per\centi\meter\cubed} and temperatures of \SI{0.1}{\electronvolt} and \SI{3.0}{\electronvolt}.
The \SI{0.1}{\electronvolt} case lies below the melting point, so the atomic configuration contains thermally broadened crystalline order, but the \SI{3.0}{\electronvolt} case is liquid.
While the higher temperature simulation is fairly well-behaved, the \SI{0.1}{\electronvolt} case supports a larger persistent current density, leading to significant sensitivity to the window function width.

The f-sum rule imposes a universal and model-independent constraint on the dynamical conductivity: 
\begin{equation}
    \int_0^\infty \mathrm{Re}[\sigma(\omega)] d\omega = \frac{\pi n_e}{2},
    \label{eq:cond_sum_rule}
\end{equation}
where $n_e$ is the average electron density \cite{kubo1957statistical}.
This rule derives from the Kramers–Kronig relations which are based on causality and thus holds for any system irrespective of the model details. 
However, similar to the challenges with the DSF sum rule (see Eq.~\eqref{eq:dsffsum}), satisfying the conductivity f-sum rule in practice is difficult with real-time TDDFT because of 
numerical limitations such as time-step and real-space grid discretizations. 

\section{Beyond linear response: stopping power}
\label{sec:stopping}

Although practically relevant response properties often fall within the linear regime, real-time simulations can also capture higher-order response contributions.
For example, real-time TDDFT calculations have predicted deviations from Ohm's law under very strong uniform electric fields \cite{andrade2018negative}.
Similarly, TDDFT can access nonlinear optical properties under high-intensity electromagnetic irradiation \cite{yabana2012time,uemoto2021first,sun2021real}.
Quadratic response functions influence resonant inelastic x-ray scattering (RIXS) spectra \cite{nascimento2021resonant}, though to our knowledge a real-time treatment has yet to be attempted.
Here, we will focus on electronic stopping power as a beyond-linear-response property 
of particular importance in HED science in the context of fusion fuel heating processes.

Real-time TDDFT can predict electronic stopping powers both within and beyond the linear-response regime by simulating the electron dynamics induced by a moving charge \cite{correa2018calculating}.
In this case, the perturbing potential takes the form
\begin{equation}
    V_{\mathrm{pert}}(\mathbf{r},t) = -\frac{Z_\mathrm{proj}}{|\mathbf{r}-\mathbf{R}_\mathrm{proj}(t)|},
\end{equation}
where $Z_\mathrm{proj}$ and $\mathbf{R}_\mathrm{proj}(t)$ are the charge and time-dependent position of the projectile, respectively.
The electronic stopping power is then given by a time average of the force exerted by the host electrons on the projectile,
\begin{equation}
    \mathbf{F}(t) = - \frac{d}{d\mathbf{R}_\mathrm{proj}} \sum_j f_j \langle \phi_j(t)| \hat{H}_\mathrm{KS}(t) | \phi_j(t)\rangle .
    \label{eq:force}
\end{equation}

The projectile is typically a positively charged ion, with significant efforts devoted to proton/deuteron and alpha-particle ($Z_\mathrm{proj}=1$ or 2) stopping in the context of ion-driven fast ignition and hot spot self-heating \cite{magyar2016stopping,ding2018ab,white2018time,kononov2024reproducibility}.
In this case, a pseudopotential may approximate $V_{\mathrm{pert}}$ in the same way as the other electron-ion interactions contributing to the external potential (see Eq.~\eqref{eq:vext} and accompanying text).
A few studies have also considered $Z_\mathrm{proj}=-1$ to model 
energetic electrons in the context of laser-target coupling in indirect-drive fusion schemes \cite{nichols2023time,nichols2024time}.

As a notable distinction from TDDFT calculations of other response properties, experimentally-relevant stopping power simulations are restricted to integer values of $Z_\mathrm{proj}$ rather than admitting arbitrarily weak perturbations.
Moreover, the exact perturbing potential diverges as $\mathbf{r}\rightarrow \mathbf{R}_\mathrm{proj}$, challenging the notion of a linear-response regime for stopping power.
In reality, the projectile's wake can modify excitations beyond linear response, and in particular, a sufficiently slow ion can capture electrons into its bound states, thus generating additional stopping power mechanisms \cite{lim2016electron,kononov2021anomalous,ullah2018core}.
Real-time TDDFT natively captures these nonlinear effects by explicitly modeling the attendant charge dynamics, but the method can also approach linear response through artificial $|Z_\mathrm{proj}|<1$ projectiles \cite{kononov_nonlinear_2025} for targeted benchmarking of other models \cite{grabowski2020review,hentschel2023improving,stanek2024review}.

\begin{figure}
    \centering
    
    \includegraphics{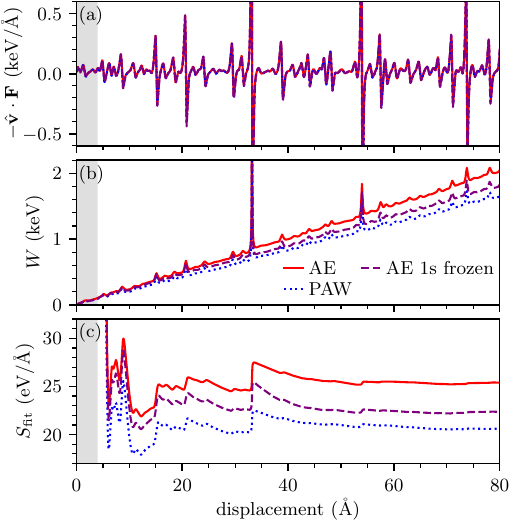}
    \caption{Exemplary TDDFT stopping power calculation.
    The friction force (a) experienced by a proton moving through warm dense carbon was integrated according to Eq.~\eqref{eq:stopping_work} to obtain the stopping work (b) in calculations that include all electrons and represent ions with bare Coulomb potentials (solid red), use bare Coulomb potentials but freeze carbon 1s states to represent an exact pseudopotential (dashed purple), or use PAW pseudopotentials and include only 4 valence electrons per carbon ion (dotted blue).
    Linear fits given by Eq.~\eqref{eq:stopping_fit} then provide the average stopping power (c) as a function of projectile displacement.
    Gray shading indicates the transient regime excluded from the data analysis.
    }
    \label{fig:stopping}
\end{figure}

Fig.~\ref{fig:stopping} shows examples of TDDFT stopping power calculations as performed in Ref.~\onlinecite{kononov_nonlinear_2025} for a proton traveling through \SI{10}{\gram\per\centi\meter\cubed} carbon at a temperature of \SI{2}{\electronvolt} with a velocity of \SI{5}{\au} and different treatments of the electron-ion interactions.
As the proton moves through the simulation cell, the friction force it experiences varies according to its local environment, including dynamically excited electrons (see Fig.~\ref{fig:stopping}a).
To derive an average stopping power, we first integrate the friction force to compute the stopping work shown in Fig.~\ref{fig:stopping}b:
\begin{equation}
    W(x) = -\int \mathbf{F}(t) \cdot d\mathbf{x(t)},
    \label{eq:stopping_work}
\end{equation}
where $\mathbf{x}(t)=\mathbf{R}_\mathrm{proj}(t)-\mathbf{R}_\mathrm{proj}(0)$ denotes the projectile's displacement and $x=|\mathbf{x}|$.
We then fit a linear model to the stopping work,
\begin{equation}
    W(x) \approx S_\mathrm{fit}\, x + W(0) \label{eq:stopping_fit},
\end{equation}
excluding transient behavior during $x<4$\SI{}{\angstrom}.
The slope $S_\mathrm{fit}$ gives the average stopping power, which converges to an approximately constant value as the proton travels through the cell and samples the environments along its path (see Fig.~\ref{fig:stopping}c).

Whereas linear-response properties generally do not depend on perturbation details, TDDFT stopping powers can be highly sensitive to the choice of $\mathbf{R}_\mathrm{proj}(t)$ \cite{schleife2015accurate,maliyov2018electronic,gu2020efficient,kononov_trajectory_2023}.
Localized electrons like core states can absorb energy and contribute to the stopping power only when the projectile passes near them \,---\, generating a strong force pulse and step-like increase in the stopping power like the features shown in Fig.~\ref{fig:stopping} \,---\, and these close encounters must be carefully sampled.
Notably, the projectile's path is essentially pre-determined through a choice of initial position $\mathbf{R}_\mathrm{proj}(0)$ and velocity vector $\mathbf{v}_{\mathrm{proj}}$.
Ehrenfest dynamics \cite{ehrenfest1927bemerkung} can capture dynamic deflection by evolving $\mathbf{v}_{\mathrm{proj}}$ according to instantaneous forces, but often the velocity would not change appreciably over the course of the \SI{}{\femto\second}-scale simulations.
Thus, maintaining a fixed velocity such that
\begin{equation}
    \mathbf{R}_\mathrm{proj}(t) = \mathbf{R}_\mathrm{proj}(0) + t \mathbf{v}_{\mathrm{proj}}
\end{equation}
usually provides a very good approximation that simplifies the selection of $\mathbf{R}_\mathrm{proj}(0)$ and $\mathbf{v}_{\mathrm{proj}}$.

To obtain practically useful average stopping powers, $\mathbf{R}_\mathrm{proj}(t)$ should be informed by experimental considerations.
A stopping power measurement \cite{malko2022proton,zylstra:2019} typically focuses an ion beam on a polycrystalline or disordered target with a finite spot size such that ions sample a distribution of different environments.
Similarly, fusion products are randomly oriented within a hot spot plasma.
Several schemes have been proposed to mimic these scenarios within a TDDFT approach.
For a crystalline material, systematic integration over a symmetry-reduced phase space of projectile trajectories is possible \cite{maliyov2018electronic}.
Alternatively, stopping power can be averaged over a random ensemble of trajectories \cite{yost2016electronic,white2022mixed}.
To ensure representative sampling of rare (but potentially large energy-transfer) close encounters with electrons localized near host ions, projectile trajectories can be optimized to reproduce a reference distribution based on the host material's atomic configuration \cite{gu2020efficient,kononov_trajectory_2023}.

The choice of $\mathbf{R}_\mathrm{proj}(t)$ can also influence finite-size effects through artificial interactions between the projectile and periodic images of its wake \cite{kononov_trajectory_2023,kononov2024reproducibility}.
As an extreme example, a particle traveling parallel to a simulation cell axis will traverse the exact same path upon crossing a periodic boundary, thus artificially interacting with pre-excited material.
This type of finite-size error can be reduced by selecting trajectories that are incommensurate with the supercell periodicity and remain relatively far from their own periodic images \cite{kononov_trajectory_2023}, averaging results over a larger ensemble of short trajectories \cite{gu2020efficient,kononov2024reproducibility}, and/or using very large supercells.
The latter is also needed to capture contributions from longer wavelength plasmonic excitations \cite{correa2018calculating} which become increasingly important for fast projectiles.

The pseudopotential approximation can have a significant, velocity-dependent effect on TDDFT stopping power predictions.
Fast projectiles are increasingly capable of exciting electrons bound to host ions, and pseudizing core states neglects these stopping power mechanisms \cite{schleife2015accurate,kononov_trajectory_2023}.
Intentionally excluding core-electron processes enables focused benchmarking of free-electron stopping power models \cite{hentschel2023improving,kononov_nonlinear_2025}.
Combining results computed with different pseudopotentials allows isolating core-electron contributions to the total stopping power, with implications for efficient convergence of finite-size errors that mainly affect the free-electron contribution \cite{kononov_trajectory_2023}.
Pseudizing core states of heavy ion projectiles can similarly restrict captured stopping power mechanisms \cite{ullah2018core}.

Pseudopotentials can also introduce more subtle errors in stopping power predictions even beyond the subset of participating electrons \cite{kononov_trajectory_2023, kononov_nonlinear_2025}.
For example, pseudopotentials bias the free-electron contribution to stopping power in Fig.~\ref{fig:stopping} relative to all-electron calculations.
Typically, pseudopotentials are designed to reproduce all-electron wavefunctions outside of a cutoff radius $r_\mathrm{cut}$ for electronic configurations close to the ground state at charge neutrality, a scenario violated by strongly excited host ions and fast projectile ions that remain highly ionized during the non-equilibrium dynamics.
A smaller $r_\mathrm{cut}$ improves transferability to situations far from equilibrium (at the expense of requiring larger basis sets), thus improving the accuracy of stopping power predictions.

Finally, an average stopping power $S_\mathrm{avg}$ can be computed from the time-dependent TDDFT data in several different ways that are consistent in principle, but can vary in terms of convergence properties.
For example, a simple time-average of the instantaneous friction force,
\begin{equation}
    S_\mathrm{avg} \approx -\frac{1}{t_1-t_0} \int_{t_0}^{t_1} \hat{\mathbf{v}}_\mathrm{proj} \cdot \mathbf{F}(t) \; dt,
\end{equation}
may suffer from sensitivity to the start and end times $t_0,t_1$, particularly in the presence of occasional strong forces from electrons localized near host ions.
Removing force contributions that contain large fluctuations but are expected to vanish or become negligible when averaged over a sufficiently long trajectory --- like ion-ion repulsion or forces induced by the equilibrium electron density --- can mitigate this sensitivity \cite{kononov2024reproducibility}.
Extracting the average stopping power as the slope of a linear fit to the stopping work (see Eqs.~\eqref{eq:stopping_work} and \eqref{eq:stopping_fit}) often provides smooth convergence to an average stopping power even without isolating non-vanishing force contributions \cite{kononov_trajectory_2023, kononov2024reproducibility}.
A similar analysis of the total electronic energy can provide equivalent results \cite{schleife2015accurate}, though small discrepancies may arise when using PAW and neglecting force correction terms arising from moving projectors \cite{ojanpera_nonadiabatic_2012,baczewski:2014}.
In any case, ignoring early transient behavior as the projectile charge state equilibrates accelerates stopping power convergence.

\section{Summary and Outlook}
\label{sec:conclusion}

Real-time TDDFT is a versatile tool for computing electronic response properties of HED systems.
Within the linear-response regime, TDDFT can predict high-fidelity dynamic structure factors and dynamic conductivities.
Real-time evolution also allows efficient access to properties beyond linear response like electronic stopping power.
The wide range of self-consistently captured physics has enabled novel theoretical insights and detailed inter-model benchmarking in regimes difficult to probe experimentally or effectively approximate theoretically.

Nonetheless, TDDFT has several limitations and multiple avenues for continued development in the context of HED physics \cite{vorberger2025roadmap}.
For one, the method inherits several challenges from static DFT, including developing XC functionals and pseudopotentials that suitably balance accuracy and efficiency across different thermodynamic conditions.
The non-equilibrium dynamics modeled by real-time TDDFT can further strain these approximations, and computationally tractable non-adiabatic XC treatments remain elusive.

As discussed in this review, numerical sensitivities can obstruct reliable TDDFT predictions in frequency extremes, currently affecting elastic x-ray scattering, DC electrical conductivity, and low-magnitude high-energy response properties at hundreds of eVs or more.
Simulation parameters like supercell size and projectile trajectory (in the case of electronic stopping power calculations) can bias TDDFT results by influencing the realistic nature of the dynamics.
Rigorously quantifying and mitigating all of these uncertainties is critical to ensure meaningful comparisons to other models, especially as TDDFT predictions continue to guide improvements to more efficient computational methods.

High computational costs limit not only prospects for systematically reducing uncertainties for existing applications, but also feasible thermodynamic regimes and accessible physical processes.
The deterministic Kohn-Sham formulation of TDDFT discussed here is generally effective for moderately ionized, degenerate systems near solid densities.
Low densities require large simulation cells, while high temperatures require more electronic states or additional approximations like stochastic or orbital-free treatments \cite{ding2018ab,sharma2023stochastic,white2020fast}.
Properties involving inner-shell electrons like x-ray opacity or the K-edge in x-ray scattering require extremely hard pseudopotentials, large basis sets, and small time steps.

Further accelerating TDDFT calculations could expand the method's range of applications and enable more widespread use within HED science.
Leveraging heterogeneous computing architectures \cite{andrade2021inq} and developing effective machine learning workflows \cite{ward2024accelerating,shah2025machine,boyer2024machine} offer two promising avenues for increasing throughput.
Extending complementary approaches that embed a subset of the electrons within an effective potential \cite{krishtal2015subsystem,gill2021time,yunus2024embedding} could also enable efficient access to core processes within real-time TDDFT.

Overall, real-time TDDFT has emerged as a valuable computational method within the WDM modeling landscape, capable of predicting accurate electronic response properties and benchmarking other models.
Exciting opportunities to further improve TDDFT's accuracy and efficiency, expand its feasibility to new regimes, and apply the method to advance HED science lie ahead.

\begin{acknowledgments}
    We are grateful to
    Alexander White,
    Alexandra Olmstead,
    Alfredo Correa,
    Amanda Dumi,
    Andr\'e Schleife,
    Brian Robinson,
    Cody Melton,
    Joshua Townsend,
    Kyle Cochrane,
    Luke Shulenburger,
    Ray Clay,
    Stephanie Hansen,
    Tadashi Ogitsu,
    and
    Xavier Andrade
    for many stimulating discussions.
    We also thank 
    Luke Stanek for generously sharing thermalized beryllium configurations,
    Joel Stevenson for HPC support,
    and Heath Hanshaw for pre-publication review.
    AK and ADB were partially supported by the US Department of Energy Science Campaign 1, and all authors were partially supported by the Sandia National Laboratories' Laboratory Directed Research and Development (LDRD) Project No.\ 233196.
    This work was performed, in part, at the Center for Integrated Nanotechnologies, an Office of Science User Facility operated for the U.S.\ Department of Energy (DOE) Office of Science.

    Sandia National Laboratories is a multi-mission laboratory managed and operated by National Technology \& Engineering Solutions of Sandia, LLC (NTESS), a wholly owned subsidiary of Honeywell International Inc., for the U.S. Department of Energy’s National Nuclear Security Administration (DOE/NNSA) under contract DE-NA0003525. This written work was authored by employees of NTESS. The employees, not NTESS, own the right, title and interest in and to the written work and are responsible for its contents. Any subjective views or opinions that might be expressed in the written work do not necessarily represent the views of the U.S. Government. The publisher acknowledges that the U.S. Government retains a non-exclusive, paid-up, irrevocable, world-wide license to publish or reproduce the published form of this written work or allow others to do so, for U.S. Government purposes. The DOE will provide public access to results of federally sponsored research in accordance with the DOE Public Access Plan. 
\end{acknowledgments}

\bibliography{main.bib}

\end{document}